

\input tables.tex

\overfullrule0pt
\normalparskip=0pt
\itemsize=19pt

\def\ls#1{_{\lower1.5pt\hbox{$\scriptstyle #1$}}}
\def\unlock{\catcode`@=11} 
\def\lock{\catcode`@=12} 
\unlock
\def\refitem#1{\r@fitem{#1.}}
\refindent=19pt
\def\refout{\par\penalty-400\vskip\chapterskip
   \spacecheck\referenceminspace
   \ifreferenceopen \Closeout\referencewrite \referenceopenfalse \fi
   \line{\hfil \bf References\hfil}\vskip\headskip
   \input \jobname.refs
   }
\def\chapter#1{\par \penalty-300 \vskip\chapterskip
   \spacecheck\chapterminspace
   \chapterreset \leftline{\bf \chapterlabel.~~#1}
   \nobreak\vskip\headskip \penalty 30000
   {\pr@tect\wlog{\string\chapter\space \chapterlabel}} }
\def\section#1{\par \ifnum\lastpenalty=30000\else
   \penalty-200\vskip\sectionskip \spacecheck\sectionminspace\fi
   \gl@bal\advance\sectionnumber by 1
   {\pr@tect
   \xdef\sectionlabel{\chapterlabel.%
       \the\sectionstyle{\the\sectionnumber}}%
   \wlog{\string\section\space \sectionlabel}}%
   \noindent {\it\sectionlabel.~~#1}\par
   \nobreak\vskip\headskip \penalty 30000 }
\lock
\sequentialequations
\def\SCIPP{\centerline {\it Santa Cruz Institute for Particle Physics}
  \centerline{\it University of California, Santa Cruz, CA 95064}}
\let\hf=\hfill
\def\rta{\rightarrow}
\def\tanb{\tan\beta}
\def\wt{\widetilde}
\def\mz{m_Z}
\def\wpm{W^{\pm}}
\def\mt{m_t}
\def\hl{h^0}
\def\mhl{m_{\hl}}
\def\ha{A^0}
\def\mha{m_{\ha}}
\def\chitil{\widetilde\chi}
\def\cnone{\chitil^0_1}
\def\cntwo{\chitil^0_2}
\def\cpone{\chitil^+_1}
\def\sq{\widetilde q}
\def\msq{M_{\tilde q}}
\def\slep{\widetilde l}
\def\snu{\widetilde \nu}
\def\gl{\widetilde g}
\def\br{BR}
\def\cpmone{\widetilde\chi^\pm_1}
\def\cmone{\widetilde\chi^-_1}
\def\mcnone{M_{\widetilde\chi^0_1}}
\def\mcpmone{M_{\widetilde\chi^\pm_1}}
\def\mslep{M_{\widetilde\ell}}
\def\msnu{M_{\widetilde\nu}}
\def\ptlmax{E_T^{\ell\,{\rm max}}}
\def\ptjmax{E_T^{j\,{\rm max}}}
\def\mgl{M_{\tilde g}}
\def\slep{\widetilde\ell}
\def\pbi{~{\rm pb}^{-1}}
%
%
\let\twocolin=Y
\newdimen\fullhsize
\newbox\leftcolumn
\def\twocol{
\def\papersize{\hsize=29.5pc \fullhsize=10.1in \vsize=6.7in
   \hoffset=-1 in \voffset=-.7 in
   \advance\hoffset by .5in \advance\voffset by .5 in
   \skip\footins=\bigskipamount \singlespace }
\Tenpoint 
\let\lr=L
\output={
    \if L\lr
      \global\setbox\leftcolumn=\columnbox
      \global\let\lr=R
    \else
      \doubleformat
      \global\let\lr=L
    \fi
    \ifnum\outputpenalty>-20000
    \else\dosupereject\fi
}
\def\fullline{\hbox to \fullhsize}
\def\doubleformat{\shipout\fullline{\box\leftcolumn\hfil\columnbox}}
\def\columnbox{\vbox{\leftline{\pagebody}\makefootline}\advancepageno}
}


\tolerance=5000
\hfuzz=2pt
\voffset=1truein

%
%
\skewchar\ninei='177
\skewchar\ninesy='60
 \def\ninebig#1{{\hbox{$\textfont0=\twelverm\textfont2=\twelvesy%
   \left#1\vbox to9.25pt{}\right.\nulldelimiterspace=0pt%
   \mathsurround=0pt$}}}
\def\ninepoint{\def\rm{\fam0\ninerm}%
  \textfont0=\ninerm \scriptfont0=\sevenrm \scriptscriptfont0=\sixrm
  \textfont1=\ninei \scriptfont1=\seveni \scriptscriptfont1=\sixi
  \textfont2=\ninesy \scriptfont2=\sevensy \scriptscriptfont2=\sixsy
  \def\it{\fam\itfam\nineit}%
  \textfont\itfam=\nineit
  \def\sl{\fam\slfam\ninesl}%
  \textfont\slfam=\ninesl
  \def\bf{\fam\bffam\ninebf}%
  \textfont\bffam=\ninebf \scriptfont\bffam=\sevenbf
   \scriptscriptfont\bffam=\sixbf
  \normalbaselineskip=11pt
  \let\sc=\ninerm
  \let\big=\ninebig
  \setbox\strutbox=\hbox{\vrule height8.0pt depth3.0pt width0pt}%
  \normalbaselines\rm}
\unlock
\footindent=20pt
\def\Vfootnote#1{\insert\footins\bgroup
   \interlinepenalty=\interfootnotelinepenalty \floatingpenalty=20000
   \singl@true\doubl@false\ninepoint
   \splittopskip=\ht\strutbox \boxmaxdepth=\dp\strutbox
   \leftskip=\footindent \rightskip=\z@skip
   \parindent=0.5\footindent \parfillskip=0pt plus 1fil
   \spaceskip=\z@skip \xspaceskip=\z@skip \footnotespecial
   \Textindent{#1}\footstrut\futurelet\next\fo@t}
\def\titlestyle#1{\par\begingroup \titleparagraphs
   \noindent #1\par\endgroup }
\def\title#1{\vskip\frontpageskip \titlestyle{\baselineskip=18pt \twelvebf #1}
\vskip\headskip }
\def\author#1{\vskip\frontpageskip\titlestyle{\twelvecp #1}\nobreak}

\def\address#1{\par\kern 5pt\titlestyle{\it #1}}
\def\andaddress{\par\kern 5pt \centerline{\sl and} \address}
\def\abstract{\par\dimen@=\prevdepth \hrule height\z@ \prevdepth=\dimen@
   \vskip\frontpageskip\centerline{\twelverm ABSTRACT}\vskip\headskip }
\lock
\papers
\Pubnum={SCIPP 93/21}
\date={July 1993}
\titlepage
\vbox to 3cm{}
\singlespace
\title{{ Phenomenology of Gluino Searches at the Tevatron}
\foot{Work supported in part by the U.S.~Department of Energy.}}
\vskip1cm
\centerline{\caps Howard E. Haber}

\SCIPP
\vskip1cm

\vbox{%
\centerline{\bf Abstract}
\vskip4pt
Present data indicates that the gluino (if it exists) must be heavier
than about 95~GeV.  During the next few years
as the Tevatron integrated
luminosity increases, gluino searches will be able
to probe the mass range between 100 and 200~GeV.  For masses in this
range, a variety of gluino decay modes can provide viable signatures
for gluino detection.  Apart from the classic missing
transverse energy signal, the detection of high transverse momentum
like-sign dileptons may be the cleanest signature for gluino production.
Other signatures such as the production of a hard photon
in the gluino cascade decay may also play an important role in
confirming the supersymmetric origin of events originating from
gluino production and decay.
}
\vskip .1in
\vfill
\centerline{Invited Talk presented at the SUSY-93 Workshop,}
\centerline{Northeastern University, Boston, MA, 29 March--1 April 1993.}
\vfill
\endpage

\chapter{The Gluino Mass Limit, circa 1992.}

The Tevatron has recently completed its 1992-93 run and has delivered
about 25 ${\rm pb}^{-1}$ of data to the CDF and
D0 detectors.  Next year, the Tevatron hopes to triple the
accumulated luminosity of the 1992-93 run.  With this large
increase of data from the highest energy collider in the world, there
is some hope that evidence for deviation from the Standard Model
could soon be at hand.  At this meeting on the status of supersymmetry
in 1993, it is appropriate to reflect on the prospects for the
discovery of supersymmetry at the Tevatron in the near future.
Since squarks and gluinos have the largest production cross-sections
at a hadron collider, it is these particles that have attracted the
major attention of the two detector Collaborations in their searches
for supersymmetry.

In supersymmetry searches at hadron colliders, two basic scenarios
emerge.  Gluinos and squarks are produced in pairs, so that possible
final states are $\wt g\wt g$, $\wt q\wt q$ and $\wt g\wt q$.%
\foot{Here, I do not distinguish between squarks and their
antiparticles.  On the other hand, gluinos are self-conjugate, a
property that plays an important role in section 4.}
If $\msq<\mgl$, then $\gl\rta\sq \bar q$ (or $\overline{\sq}q$), and
one should first concentrate on the signatures of squark production at the
Tevatron.  Likewise, if $\mgl<\msq$, then $\sq\rta q\gl$, and it is
appropriate to first focus attention on gluino production.
Supersymmetric models do not yield a definitive statement as to
which case is more likely to be realized in nature, although there may be
a slight theoretical bias in favor of the lighter gluino.  In this
paper, I choose to consider the case of $\mgl<\msq$ and examine the
phenomenology of gluino production and decay at the Tevatron.
\REF\pdg{K. Hikasa \etal\ [Particle Data Group], {\it Phys. Rev.}
{\bf D45} 1 June 1992, Part II, pp.~IX.5--IX.12.}

\TABLE\glustatus{}
During the past decade, many experiments have conducted searches for
gluinos.\refmark\pdg\
The results of all such searches have been negative.
{}From these negative searches, gluino mass limits have been obtained.
In the literature, there has been some disagreement on the extracted
gluino mass limits.
A comprehensive discussion of these disagreements lies beyond the
scope of this paper.  Instead,
I will state my personal opinions, and indicate how I draw
my conclusions. A brief summary on the status of gluino searches is
presented in Table \glustatus.

\REF\haberkane{For a summary of the relevant data, see H.E. Haber and
G.L. Kane, {\sl Phys. Rep.} {\bf 117} (1985) 75.  For more recent
data, see, \eg\ T. Akesson \etal\ [HELIOS Collaboration],
{\sl Z. Phys.} {\bf C52} (1991) 219.}
\REF\jlf{J. Lee-Franzini, in {\it Supernova 1987A, One Year Later:
Results and Perspectives in Particle Physics},
Proceedings of Les Rencontres de Physique de la Valle d'Aosta, La
Thuile, Italy, February 26---March 5, 1988, edited by M. Greco
(Editions Frontieres, Gif-sur-Yvette, France, 1988) p.~301.}
\REF\bhk{R.M. Barnett, H.E. Haber, and G.L. Kane, {\sl Nucl. Phys.}
{\bf B267} (1986) 625.}
\REF\uai{C. Albajar \etal\ [UA1 Collaboration], {\sl Phys. Lett.}
{\bf B198} (1987) 261.}
\REF\uaii{J. Alitti \etal\ [UA2 Collaboration], {\sl Phys. Lett.}
{\bf B235} (1990) 363.}
\REF\cdf{F. Abe \etal\ [CDF Collaboration], {\sl Phys. Rev. Lett.}
{\bf 69} (1992) 3439.}
\REF\hidaka{K. Hidaka, {\sl Phys. Lett.} {\bf D44} (1991) 927.}

\topinsert
\def\tg{\widetilde g}
\centerline{\bf Table \glustatus. Gluino Mass Limits}
\vskip4pt
\ninepoint \baselineskip=11pt
\centerline{\vbox{%
\halign to \hsize{%
#\hfil&\tabskip1em plus 1fil
\vtop{ \hsize=3.56cm \parindent=0pt \raggedright\strut #\strut}&
\vtop{ \hsize=4.66cm \parindent=0pt \raggedright\strut #\strut}%
\tabskip=0pt \cr
\noalign{\vglue2.3pt\hrule\vskip2pt\hrule\vglue2.3pt}
\hf $M_{\tg}$ Sensitivity\quad\hf& \hf Experiment \hf& \hf Comments \hf\cr
\noalign{\vglue2.3pt\hrule\vglue4.25pt}
$M_{\tg}\lsim 1$--10 GeV& A variety of beam dump experiments and beam
                     contamination searches\refmark\haberkane\ &
           Neutron backgrounds are problematical if $\mgl\simeq m_n$;
           mass limits depend on gluino lifetime \cr
\noalign{\vskip2.3pt\hrule\vskip5pt}
$M_{\tg}\lsim 1$ GeV& $\psi \to \gamma\eta_{\tilde g}$&
$\eta_{\tilde g} = 0^{-+}\
    \tg\tg$ bound state; \cr
\noalign{\vskip2.3pt}
$M_{\tg}\lsim 3$ GeV&  $\Upsilon \to \gamma\eta_{\tilde g}$\refmark\jlf\
& color factor of 27/4
     enhancement over production of $\eta_Q$ \cr
\noalign{\vskip2.3pt\hrule\vskip4pt}
$3\lsim M_{\tg} \lsim 4$ GeV& Based on UA1 data&  Analysis
      of ref.~\bhk; important use of $\tg\tg g$ production
\cr \noalign{\vskip2.3pt\hrule\vskip4pt}
$4\lsim M_{\tg} \lsim 53$ GeV& UA1\refmark\uai\  & Assumes $\tg\to
       q\bar q\widetilde\gamma$ with  \cr
$16 \lsim M_{\tg} \lsim 79$ GeV\qquad&  UA2\refmark\uaii\ &
         $M_{\tilde\gamma} \lsim 20$ GeV\cr
\noalign{\vskip2.3pt\hrule\vskip4pt}
$30 \lsim M_{\tg} \lsim 95$ GeV&  CDF\refmark\cdf\ &  Upper limit
        includes effects of gluino cascade decays with \nextline
        $\mu=-250$~GeV  and $\tanb=2$\cr
\noalign{\vskip2.3pt\hrule\vskip4pt}
$M_{\tg}\lsim 132$ GeV&  ref.~\hidaka\  &  Based on CDF gluino mass limits
        and LEP limits on chargino and neutralino masses [assuming the
        validity of eq.~(2.1)].\cr
\noalign{\vskip2.3pt\hrule\vskip2pt\hrule}
\cr
} } }
\endinsert

\REF\lightgluino{L. Clavelli, {\sl Phys. Rev.} {\bf D45} (1992) 3276;
{\bf D46} (1992) 2112; M. Jezabek and J.H. K\"uhn, {\sl Phys. Lett.}
{\bf B301} (1993) 121; J. Ellis, D.V. Nanopoulos, and D.A. Ross,
{\sl Phys. Lett.} {\bf B305} (1993) 375.}
Two main questions have been debated: (i) are very
light gluinos allowed? and (ii) what is the maximum gluino mass ruled
out by collider data?
It is convenient to separate the discussion into two
mass regimes: $\mgl<3$~GeV and $\mgl>3$~GeV.  The borderline between
these two regions has been called the ``light gluino window'' in the
literature.  Recently, there has been a revival of interest in this
region.  A light gluino has been advocated in order to explain a
possible inconsistency in the determination of $\alpha_s$ by
different techniques.\refmark\lightgluino\
Of course, experimental searches will be the
final arbiter of the existence or non-existence of the light gluino
window.

\REF\goldman{J.H. K\"uhn and S. Ono, {\sl Phys. Lett.} {\bf 142B}
(1984) 436; W.-Y. Keung and A. Khare, {\sl Phys. Rev.} {\bf D29}
(1984) 2657; T. Goldman and H.E. Haber, {\sl Physica} {\bf D15}
(1985) 181.}
\REF\cusb{P.M. Tuts \etal\ [CUSB Collaboration], {\sl Phys. Lett.}
{\bf B186} (1987) 233.}
A variety of beam dump experiments and
beam contamination searches rule out gluinos with $\mgl<3$~GeV,
with some dependence on the gluino lifetime.\refmark\haberkane\
Perhaps the most
definitive negative result is the non-existence of a pseudoscalar
$\gl\gl$ bound state (sometimes called the $\eta_{\tilde g}$).
Such a state, if it existed, would be prominent in radiative
quarkonium decay.\refmark\goldman\
Because gluinos are color octets, the
decay rates for $\psi\rta\gamma\eta_{\tilde g}$ and $\Upsilon\rta
\gamma\eta_{\tilde g}$ are enhanced by a color factor of $27/4$
over the radiative decay into $\gamma\eta_Q$, where $\eta_Q$ is a
pseudoscalar meson bound state of color-triplet quarks (\eg, $\eta$,
$\eta^\prime$, and $\eta_c$).  No such prominent
pseudoscalar state is seen in the data.   The published limits
of the CUSB collaboration\refmark\cusb\
at CESR rules out gluino masses between
0.6~GeV and 2.2~GeV.  Subsequent analysis of the CUSB data\refmark\jlf\
extended the upper mass limit to 2.6~GeV.
Presumably, additional $\Upsilon$ decay data now exists that could push
the gluino mass bound above 3~GeV.

For gluino masses above 3~GeV, one must turn to the collider data.
Gluino mass limits have been presented by the UA1\refmark\uai\
and UA2\refmark\uaii\
Collaborations based on CERN $p\overline p$ Collider data and by the
CDF Collaboration\refmark\cdf\
based on Tevatron data.  These analyses assume that
the gluino decays inside the detector with the emission of the
lightest supersymmetric particle (LSP) which carries off undetected
(``missing'') transverse energy.  Only the UA1 data is sensitive to
light gluino masses.  The UA1 limits rule out gluinos in the range
$4\leq\mgl\leq 53$~GeV.  This apparently leaves open the light gluino
window.  However, I believe that the UA1 limits are too conservative.
In particular, the UA1 analysis did not take into account the
mechanism $gg\rta\gl\gl g$ which is especially important for $\mgl<5$~GeV.
In particular, the kinematical configuration in which a hard gluon
recoils against the two gluinos which are emitted in the same
hemisphere can lead to substantial missing energy and pass the UA1
cuts.  Barnett, Kane and I investigated this mechanism
carefully\refmark\bhk\ and concluded that the UA1 data definitively
ruled out gluino masses as light as 3~GeV.  This essentially
eliminates the possibility of the light gluino window.

The analyses of the UA2 and CDF Collaborations are not sensitive to
the light gluino.  However, their excluded mass regions overlap those
of the UA1 analysis and therefore extend the UA1 gluino mass bound.
The UA2
Collaboration concluded that $\mgl<79$~GeV.  The CDF Collaboration
has extended the gluino mass bound further; published
results\refmark\cdf\ have been
presented based on 4.3~pb$^{-1}$ of data from the 1989-91 Tevatron
runs. In order to properly interpret the CDF data and extract gluino
mass limits, one must incorporate all the allowed gluino decay
patterns in the analysis.   Typically, one assumes
that the LSP is the lightest neutralino~%
($\cnone$).  Then, the simplest gluino decay is $\gl\rta q\bar q\cnone$,
sometimes called ``the direct decay to the LSP''.  This
is expected to be the
dominant decay for light gluinos.\foot{There is an implicit
assumption here that the gluino is heavier than $\cnone$.  Otherwise,
the gluino would be the LSP and hence stable.
Note that the CUSB limits quoted above also apply to stable gluinos.
Beam contamination experiments then rule out stable gluinos for
$\mgl<10$~GeV.\refmark\haberkane\  Heavier
stable gluinos are disfavored for both theoretical and cosmological
reasons, although this possibility cannot be completely excluded.}
For heavier gluinos, decays into heavier neutralinos and
charginos
$$\eqalign{&\gl\rta q\bar q\widetilde\chi_i^0\cr
&\gl\rta q\bar q^\prime \widetilde\chi^\pm_i\cr}\eqn\gldecays$$
become kinematically allowed, in which case, the branching ratio for
the direct decay to the LSP drops significantly below 1.  This is
relevant for phenomenology since it is the direct decay into the LSP
that produces the largest missing energy signature.  When the gluino
decays first into heavier neutralinos or charginos, these particles
subsequently
decay, eventually producing the LSP at the end of the
decay chain.  The end result is a gluino cascade decay that produces
less missing energy than the direct decay to the LSP.

\REF\tasi{For a complete description of the parameters of the MSSM,
see, \eg, H.E. Haber, SCIPP-92/33 (1993), to appear in the
Proceedings of the 1992 Theoretical Advanced Study Institute,
Boulder, CO, June 1992.}
The CDF Collaboration has presented\refmark\cdf\
gluino mass limits under the
assumption that $BR(\gl\rta q\bar q\cnone)=1$.  However, such mass
limits are not very meaningful, since they correspond to a very
unlikely choice of parameters for the supersymmetric model.
For a more realistic example, consider
the minimal supersymmetric extension of the Standard Model (MSSM),
with gaugino Majorana mass parameters related according to the
unification relation
$$M_i/M_j=\alpha_i/\alpha_j\,,\eqn\gaugino$$
where the $i$ labels the gauge group, $\alpha_i\equiv g_i^2/4\pi$,
and $g_1^2\equiv (5/3)g^{\prime2}$.  In this case, the neutralino and
chargino masses and mixing angles are determined by three parameters:
the gluino mass
($\mgl$), the ratio of Higgs vacuum expectation values ($\tan\beta$)
and the supersymmetric Higgs mass parameter ($\mu$).\refmark\tasi\
Thus, for a given value of gluino mass, the parameters $\mu$ and
$\tan\beta$ determine the relative branching ratios of the gluino
into charginos and neutralinos [eq.~\gldecays].  For gluino masses
now being probed at the Tevatron, the probability of the cascade
decays (in which the LSP is produced at the end of a multistep
decay chain) is typically
larger than the probability of
direct decay to the LSP.  Thus the missing energy
signal is degraded and the true CDF gluino mass limit should be less
than the quoted limit for $BR(\gl\rta q\bar q\cnone)=1$.
In ref.~\cdf, the CDF
Collaboration has included the effect of the cascade decays on their
analysis.  For one typical set of parameters, they deduced an upper
gluino mass bound of about 95~GeV.  This bound will surely be
improved (by both the CDF and D0 Collaborations)
once the 1992-93 Tevatron data is analyzed.

The bound $\mgl>95$~GeV is the best bound presently available based on
direct gluino searches.  However, under the assumption of
the unification of gaugino mass parameters [eq.~\gaugino], the search
for neutralinos and charginos at LEP can indirectly lead to a bound
on $\mgl$.  For example, in ref.~\hidaka, Hidaka concludes that
$\mgl>132$~GeV, based on the LEP limits on neutralino and chargino
masses and including the effects of cascade decays on the CDF gluino
search analysis mentioned above.\foot{The inferred gluino mass limit of
ref.~\hidaka\ was based in part on a preliminary CDF analysis of their
1989--91 data, and should probably be reduced slightly in light of
the published CDF results.  Note that the LEP limits
on chargino and neutralino masses are basically at their kinematic
limits and hence will not change with further running at
$\sqrt{s}=m_Z$.}

To conclude this introduction, my personal opinion is that based on
current published data, $\mgl\gsim 125$~GeV, with no open light gluino
window.  Looking to the near future, the Tevatron runs in 1992-93 and
1993-94 will yield about 25~pb$^{-1}$ and 75~pb$^{-1}$ per detector,
and should increase the gluino mass sensitivity to about 200~GeV.
The purpose of this talk is to indicate some of the phenomenological
methods available to achieve this sensitivity.  In section 2, I
present a brief review of the theory of gluino cascade decays.  For
gluino masses that can be detected at the Tevatron, there are three
important gluino decays: (i) the direct decay to the LSP ($\gl\rta
q\bar  q\cnone$), (ii) $\gl\rta q\bar q\cntwo$, and (iii) $\gl\rta q\bar
q^\prime \widetilde\chi^\pm_1$.  In case (ii), a viable signature
results if $\cntwo\rta\cnone\gamma$; this is the subject of section 3.  In
case (iii), note that the Majorana nature of the gluino%
\foot{Here, I am using a broader definition of Majorana
to include neutral particles that transform under real representations
of the underlying Standard Model gauge group.}
leads to
equal probability for producing either sign chargino.  Thus, if the
chargino decays semi-leptonically, $\gl\gl$ events can lead to like-%
sign dileptons in the final state.  This is the subject of section 4.

\chapter{The Theory of Gluino Cascade Decays}

\REF\xsections{
P.R. Harrison and C.H. Llewellyn Smith
{\it Nucl. Phys.} {\bf B213} (1983) 223 [E: {\bf B223} (1983) 542];
S. Dawson, E. Eichten and C. Quigg, {\it Phys. Rev.}
{\bf D31} (1985) 1581.}
At the Tevatron, the dominant production mechanism for gluinos is
$gg\rta\gl\gl$ via $s$-channel gluon and $t$- and $u$-channel gluino
exchange.\refmark\xsections\
The $\gl\gl g$ coupling follows from QCD gauge invariance,
so that the gluino cross-section
depends on only one unknown parameter---the gluino mass.  Gluinos can
also be produced in $q\bar q$ annihilation via squark-exchange.
However, this produces only a minor change in the overall gluino
cross-section.  Thus, gluino phenomenology depends on the gluino mass
and the parameters that govern the gluino decay branching ratios.
The most important gluino decay modes are tree-level three body
decays into a neutralino or chargino and a $q\bar q$ pair
[eq.~\gldecays].  With the exception of $\gl\rta q\bar q\cnone$, the
neutralino or chargino in the final state will decay into a lighter
neutralino or chargino and so on until the LSP (assumed to be $\cnone$)
is produced.  Branching ratios for the gluino cascade decays depend on
various supersymmetric parameters.  For convenience, I list below the
parameters of the MSSM:\refmark\tasi\
\vskip2.3pt
\pointbegin
The gaugino Majorana mass parameters ($M_1$, $M_2$ and $M_3$).
\point
The supersymmetric Higgs mass parameter, $\mu$.
\point
The ratio of Higgs vacuum expectation values, $\tan\beta$.
\point
The mass of the CP-odd Higgs scalar, $\mha$.
\point
Diagonal ($\widetilde f_L$-$\widetilde f_L$ and $\widetilde f_R$-%
$\widetilde f_R$) squark and slepton mass parameters.
\point Off-diagonal ($\widetilde f_L$-$\widetilde f_R$) squark and
slepton mass parameters (which depend on $\mu$, $\tanb$ and various
``$A$''-parameters.)
\vskip2.3pt

\noindent The gluino mass is given by $\mgl=|M_3|$, and
the chargino and neutralino
masses and mixing angles are determined by $M_1$, $M_2$, $\mu$ and
$\tanb$.  The Higgs sector parameters are fixed by $\tanb$ and
$\mha$.  This is relevant to the considerations here since
neutralinos and charginos can decay into final states containing
Higgs bosons.  Finally, gluino production rates and branching ratios
depend very weakly on squark masses.  (Here, I assume that $\mgl<\msq$
and $\gl\sq$ and $\sq\sq$ production rates can be neglected.)
As mentioned above, the $\gl\gl$ cross-section is dominated by
gluon and gluino exchange.  The gluino decays via
squark exchange, but this just means that the squark mass determines
the overall normalization of the gluino decay width.  To the extent
that squarks are roughly degenerate in mass, gluino {\it branching
ratios} will be independent of the squark mass.\foot{The
coupling of gluinos to $q\widetilde q$ is independent of the flavor
of $q$.  Moreover, with the exception of the top-squark, all squarks
are expected to be roughly degenerate in mass.  Interesting effects
could arise if top-squark mixing effects are important.  However, the
decay $\gl \rta t\bar t\cnone$ is not kinematically allowed for gluinos
accessible to Tevatron searches.  Thus, it is probably safe to ignore
the variation of squark masses.}  For heavy gluinos relevant for
Tevatron searches, the gluino decays essentially instantaneously
(with no visible gap between production and subsequent decay).
In considering the subsequent decay
of charginos and neutralinos in the gluino decay chain, all the
supersymmetric parameters enter to some extent.  For example,
charginos and neutralinos produced in gluino decays at the Tevatron will
typically decay via
squark, slepton, $W$ or $Z$ exchange into three-body final states.
However, even in this case,
the sensitivity to squark and slepton parameters is not
significant, since squark and slepton masses almost certainly lie
above $\mz$.

Two critical assumptions underlie the analysis presented in this
paper.  First, as discussed in section 1,
I shall assume that the gluino is lighter than all
squark masses (with the possible exception of the top-squark).
As noted above, once I assume that
$\mgl<M_{\tilde q}$, the precise value of $M_{\tilde q}$ is
not important.
Second, I shall impose the grand unification relation which states
that the gaugino mass parameters $M_1$, $M_2$ and $M_3$ are all equal
at some very large grand unification (or Planck) scale.  I then use
one-loop renormalization group evolution
to determine the value of the $M_i$ at the
electroweak scale.  One finds that the scaling of
$M_i$ is proportional to
the squared coupling constants $\alpha_i$ [eq.~\gaugino].  As a
result, the low-energy values of $M_1$ and $M_2$ can be expressed in
terms of the gluino mass
$$M_2=(g^2/g_s^2)\mgl\simeq 0.285\mgl\,,\qquad\qquad
M_1=(5g^{\prime2}/3g^2)M_2\simeq 0.483M_2\,.\eqn\glguts$$
This relation underlies nearly all phenomenological analyses of
the MSSM that appear in the literature.
Perhaps it is time to begin to consider some of the
phenomenological consequences of the violation of eq.~\glguts.
However, I will not pursue this alternative here.

\REF\rchi{R.M. Barnett, J.F. Gunion, and H.E. Haber, {\sl Phys. Rev.
Lett.} {\bf 60} (1988) 401.}
\REF\bgh{R.M.~Barnett, J.F.~Gunion and H.E.~Haber, {\sl Phys. Rev.}
{\bf D37} (1988) 1892.}
\REF\ghino{J.F. Gunion and H.E. Haber, {\sl Phys. Rev.} {\bf D37} (1988)
2515.}
\REF\extra{G. Gamberini, {\sl Z. Phys.} {\bf C30} (1986) 605;
H. Baer, V. Barger, D. Karatas, and X. Tata, {\sl Phys. Rev.}
{\bf D36} (1987) 96.}
Barnett, Gunion and I have studied
the branching ratios for the gluino, neutralino and
chargino as a function of the supersymmetric
parameters.\refmark{\rchi-\ghino}
(See also the results of ref.~\extra.)  Below, I
summarize some of the highlights of our investigation.

\FIG\bghi{The branching ratios for $\gl\rta q\bar q\widetilde\chi^0_i$
and $\gl\rta q\bar q^\prime\widetilde\chi^\pm_j$ as a function
of $\mu$ for $\mgl=120$ and 300~GeV and $\tanb=1.5$ and 4.  Sections
of the curves that are not plotted correspond to parameter choices that
yield $\mcpmone<\mz/2$. Taken from Ref.~\bgh.}
\vskip2.3pt
\pointbegin
The gluino branching ratio for the direct decay to the LSP is small
over a substantial portion of parameter space.  This conclusion
relies on the relation among the gaugino mass parameters
[eq.~\glguts].  In particular, $BR(\gl\rta q\bar q\cnone)\leq 0.14$ for
$\mgl\gsim 500$~GeV.  For gluino masses accessible to future Tevatron
searches, this branching ratio may be somewhat larger (depending on the
values of $\mu$ and $\tanb$), although it is
less than 50\% in all but a narrow region of parameter space, as
shown in fig.~\bghi.  Thus,
most gluino decays will be cascade decays, which dilute the famous
missing transverse energy signature.

\point
The dominant gluino decay mode [except for regions of parameter space
where $\tanb$ is near 1 and $|\mu|\lsim{\cal O}(\mz)$]
is $\gl\rta\cpmone+X$ where $X$ is either $u\bar d$ or $c\bar s$
(or the charge conjugate pair, depending on the sign of $\cpmone$).
For example, if $\tanb=4$
and $\mgl=120$~GeV, the sum of the branching ratios for gluino decay
into $\cpmone$ is approximately 58\% (with very weak dependence on
$\mu$; see fig.~\bghi).
Moreover, because the gluino is a Majorana fermion,
$$\Gamma(\gl\rta \bar ud\cpone)=\Gamma(\gl\rta u\bar
d\cmone)\,.\eqn\majorana$$
Thus, in $\gl\gl$ production, like-sign charginos can be produced,
which can lead to like-sign dileptons if both charginos decay semi-%
leptonically.  This is the subject of section 4.

\point
The decay $\gl\rta q\bar q\cntwo$ is competitive with the direct decay
into the LSP.  It is the second most important gluino decay mode over
a substantial range of MSSM parameter space.

\REF\previouscc{
H. Baer, X. Tata, and J. Woodside, {\sl Phys. Rev.} {\bf D41}
(1990) 906; {\bf D45} (1992) 142.}
\REF\lephiggs{See, \eg, D. Decamp \etal\ [ALEPH Collaboration],
{\sl Phys. Rep.} {\bf 216} (1992) 253.}
\REF\gordy{For a review and guide to the literature, see:
H.E. Haber, in {\it Perspectives on Higgs Physics}, edited
by G.L. Kane (World Scientific, Singapore, 1993) p.~79.}
\REF\nilles{H.-P. Nilles, {\sl Phys. Rep.} {\bf 110} (1984) 1.}
\vskip4pt
The phenomenology of gluino cascade decays\refmark\previouscc\
will depend in detail on
the neutralino and chargino branching ratios.  Since we are
interested here in gluinos in the mass range between 100 and 200 GeV,
the corresponding masses of $\cpmone$ and $\cntwo$ will be such that
tree-level two body decays $\cpmone\rta W^\pm\cnone$ and $\cntwo\rta\cnone Z$
are kinematically forbidden.  If no other tree-level two body
decays are possible, then the dominant decays of $\cpmone$ and $\cntwo$
will be: $\cpmone\rta\cnone f\bar f^{\prime}$ and $\cntwo\rta\cnone f\bar f$,
where $f$ is either a quark or lepton.  Although this is certainly
the most likely situation, it is important to consider two alternative
cases.  If neutral Higgs bosons (the CP-even $\hl$ and the CP-odd
$\ha$) are sufficiently light, then the two-body decays
$\cntwo\rta\cnone\hl$ and $\cntwo\rta\cnone\ha$ could be kinematically
allowed.  For example, ref.~\ghino\ argued that a significant region
of parameter space exists in which $\cntwo\rta\cnone\hl$ is the dominant
decay mode.  Since that analysis, two new pieces of information
suggest that the Higgs decay mode of $\cntwo$ will not be significant
at the Tevatron gluino search.  First, the LEP Higgs search implies
that $\mhl\gsim 60$~GeV.\refmark\lephiggs\
Second, radiative corrections can significantly
increase the theoretical expectation for $\mhl$.\refmark\gordy\
As a result, the
``3-body decay region'' discussed in ref.~\ghino\ expands
substantially at the expense of the ``light-Higgs region''.  Finally,
although I have assumed in this paper that squarks are heavier than
gluinos, it could happen that sleptons are significantly lighter than
the squarks (as suggested by renormalization group evolution
of low-energy supergravity models\refmark\nilles).  If
the sleptons are lighter than the lightest chargino or neutralino,
then two-body decays such as $\cpmone\rta\ell\widetilde\nu$,
$\cpmone\rta\widetilde\ell\nu$, $\cntwo\rta\ell\widetilde \ell$, and/or
$\cntwo\rta\nu\widetilde\nu$ may be kinematically allowed.  We shall
see in section 4 that if these two-body chargino decay modes are
present, then the like-sign dilepton signal in
$\gl\gl$ events is significantly enhanced.

\REF\wyler{H.E. Haber and D. Wyler, {\sl Nucl. Phys.} {\bf B323}
(1989) 267.}
Finally, there is one more two-body decay mode of potential
significance---the radiative decay $\cntwo\rta\cnone\gamma$.\refmark\wyler\
This is not
a tree-level process; it occurs only at one-loop.  Nevertheless, it
turns out that the radiative decay mode can compete with the three-%
body modes (and in some cases it can be the dominant mode) in certain
parameter regimes.  This is the subject of section 3.

\REF\bbrr{H. Baer and E. Berger, {\sl Phys.~Rev.} {\bf D34} (1986)
1361 [E: {\bf D35} (1987) 406];
E. Reya and D.P. Roy, {\sl Z. Phys.} {\bf C32} (1986) 615.}
We have seen above that the branching ratio for $\gl\rta q\bar q\cnone$
is expected to be small (less than 50\%, and more likely closer to
about 20\%).  As a result, the missing transverse energy signal is
degraded relative to the case of a 100\% branching ratio for the
direct decay to the LSP.  Nevertheless, the missing transverse
energy signature will still be important.\refmark\bbrr\
For example, $\gl\gl$
events in which one $\gl$ decays directly to the LSP and the second
$\gl$ decays arbitrarily will still  produce a significant missing
transverse energy signal.  However, in order to maximize the efficiency of
the Tevatron gluino search, other gluino signatures should be
examined.  In particular, we noted above that for $100\lsim\mgl \lsim
200$~GeV,
$$BR(\gl\rta\cpmone q\bar q^{\prime})>BR(\gl\rta\cntwo q\bar q)
>BR(\gl\rta\cnone q\bar q)\eqn\brorder$$
over nearly all of the relevant MSSM parameter space.  Thus, it is
important to examine distinctive gluino signatures arising from
the $\gl$ decay to $\cpmone$ and $\cntwo$.

\chapter{Hard Photons in Gluino Cascade Decays}

\REF\mikamo{S. Mikamo, private communication.}
\REF\preprint{H.E. Haber, preprint in preparation.}
Consider $\gl\gl$ events in which one gluino decays to $\cntwo$ via
$\gl\rta q\bar q\cntwo$, while the decay mode of the second gluino is
arbitrary.  In this section, I shall consider the case where
$\cntwo\rta\cnone\gamma$.  Such events are characterized by the emission
of a hard photon, one or more
large $E_T$ jets, and significant missing $E_T$.  Events of this
type are already being searched for at the Tevatron.\refmark\mikamo\
In this section, I shall present the preliminary results of a
calculation to estimate the fraction of $\gl\gl$ events with a hard
photon.\refmark\preprint\

\FIG\cnradone{$BR(\cntwo\rta\cnone\gamma)$ as a function of the MSSM
parameter $\mu$, for $\tanb=2$.
The three curves correspond to different choices of
gluino masses as indicated on the figure.}
\FIG\cnradtwo{$BR(\cntwo\rta\cnone\gamma)$ as a function of the MSSM
parameter $\mu$, for $\tanb=10$.
The three curves correspond to different choices of
gluino masses as indicated on the figure.}
\FIG\cnradthree{$BR(\cntwo\rta\cnone\gamma)$ as a function of $\tanb$
for two different sets of gluino masses and choices of $\mu$.}

The branching ratio for gluino decay into
$\cntwo$ is typically between 20\% and 30\% for gluino masses of
interest to the Tevatron search.  We then need to compute
$BR(\cntwo\rta\cnone\gamma)$ as a function of the MSSM parameters.
The complete
amplitude for the one-loop process $\cntwo\rta\cnone\gamma$ has been
computed in ref.~\wyler.  To evaluate the branching ratio, one must
also compute the sum of all tree-level decays of the $\cntwo$.  It is
clear that any kinematically allowed tree-level $\cntwo$
decay into two-body final states would dominate the one-loop
radiative decay (as well as tree-level decays into three-body final
states).  For the range of neutralino masses
relevant for the Tevatron gluino search, the
only possible allowed two-body (tree-level)
decays are $\cntwo\rta\cnone\hl$ and
$\cntwo\rta\cnone\ha$.  However, as argued in section 2, if
one takes into account
the LEP Higgs mass limits and the results of the
 one-loop corrected MSSM Higgs masses, one would conclude that
the tree-level $\cntwo$ decays to $\hl$ and/or $\ha$ are likely to
be present only over a small
region of parameter space.  Henceforth, I shall assume that these
decays are kinematically forbidden.  In this case, the three-body
(tree-level)
$\cntwo$ decay modes are dominant.  However,
I shall now show that the radiative decay
$\cntwo\rta\cnone\gamma$ is competitive over a rather large region of
parameter space of interest to the Tevatron gluino search.

In figs.~\cnradone--\cnradthree, I have plotted
$BR(\cntwo\rta\cnone\gamma)$ as a function of MSSM parameters for a
variety of gluino masses.  These figures show that in the region of
negative $\mu$,\foot{The LEP limits on neutralino and chargino masses
tend to rule out regions of $-40~{\rm GeV}\lsim\mu\lsim 80~{\rm GeV}$ (with
some variation depending on $M_2$ and $\tanb$).\refmark\lephiggs}
the typical value of
$BR(\cntwo\rta\cnone\gamma)$ is 10\%.  This means that in the Tevatron
gluino search, approximately 5\% of all $\gl\gl$ events should
contain a hard photon originating from $\cntwo\rta\cnone\gamma$.  It
remains to be seen whether the reducible backgrounds can be
removed (\eg, events in which a neutral jet or leading $\pi^0$ is
misidentified as a photon) thereby leaving a viable signal.  I
suspect that the gluino will not be discovered via the radiative
$\cntwo$ decay.  However, if evidence for the gluino is uncovered in
other channels, the detection of hard photon events could help to
confirm the gluino interpretation of the other signals as well as
pinpoint some of the MSSM parameters.
\vskip-6pt
\chapter{Discovering Gluinos with Like-Sign Di-Leptons}

\REF\likesign{R.M. Barnett, J.F. Gunion and H.E. Haber, LBL-34106 (1993).}
\REF\previousa{
R.M. Barnett, J.F. Gunion, H.E. Haber,
in {\it Proc. of 1988 Summer Study on High Energy Physics in the
1990's, Snowmass, Colorado, June 27-July 15, 1988}, edited by S. Jensen
(World Scientific, Singapore, 1989), p. 230; and
in {\it ``Research Directions for the Decade,'' Proc. of the 1990
Summer Study on High Energy Physics, June 25 - July 13, 1990, Snowmass, CO},
edited by E. L. Berger (World Scientific, Singapore, 1992), p. 201.
}
\REF\bkp{V. Barger, W.-Y. Keung and R.J.N. Phillips, {\sl Phys.~%
Rev.~Lett.} {\bf 55} (1985) 166.
}
\REF\previousc{
H. Baer, X. Tata, and J. Woodside,
in {\it Proc. of 1988 Summer Study on High Energy Physics in the
1990's, Snowmass, Colorado, June 27-July 15, 1988}, edited by S. Jensen
(World Scientific, Singapore, 1989), p. 220.
}
\REF\stange{V. Barger, A.L. Stange, and R.J.N. Phillips, {\sl Phys. Rev.}
{\bf D45} (1992) 1484.}
\REF\bkt{H. Baer, C. Kao and X. Tata, FSU-HEP-930527 (1993).}
\REF\previousdd{
Solenoidal Detector Collaboration Technical Design Report,
E.L. Berger \etal, Report SDC-92-201, SSCL-SR-1215, 1992, p. 3-59;
GEM Technical Design Report, W.C. Lefmann \etal, Report GEM-TN-93-262,
1992, p. 2-55.
}
\REF\previousd{
R.M. Barnett, to appear in {\it Proceedings of the 23rd
Workshop of the INFN Eloisatron Project ``Properties of SUSY Particles'',
September 28 - October 4, 1992, Erice, Italy}, (report LBL-33422).
}

In this section, I shall focus on the striking experimental
signature of two isolated leptons which can arise
{}from gluino pair production. Half of the events of this type will have
leptons with the same sign of electric charge.
This signature, which is analogous to the opposite-sign lepton signal for
the top quark, may yield sensitivity much superior to the missing-energy
gluino signature at the Tevatron, depending on the parameters of the
supersymmetric model.  The results of this section are based on work in
collaboration with R.M. Barnett and J.F. Gunion\refmark\likesign\
(see also ref.~\previousa\ for an earlier version of this
work).  Complementary work that
also considered the like-sign dilepton signature can be found in
refs.~\previouscc\ and \bkp--\previousd.

As discussed in section 2, for gluino masses accessible to the Tevatron
search, the dominant gluino decay mode is $\gl\rta q\bar q \cpmone$,
over a large range of supersymmetric parameter space.
Leptons can result from decays such as
$\cpmone\to W^{\pm}\cnone\to \ell^\pm \nu\cnone$, where the $W^{\pm}$
is either on-shell (if kinematics allow) or off-shell.  If the
sleptons are lighter than the chargino, then the two-body decays
$\cpmone\to \slep^\pm\nu \to \ell^\pm \nu\cnone$
and $\cpmone\to \widetilde\nu \ell^\pm \to \ell^\pm \nu\cnone$ are allowed
and may dominate.  Since the gluino is a Majorana fermion,
it has the distinctive property of decaying with equal
probability into fermions and antifermions.
Thus, an excellent signature for pair production of gluinos
results from events in which both gluinos decay to a chargino of the same
sign, yielding {\it like-sign} dileptons ($\ell^+\ell^+$ or $\ell^-\ell^-$)
in the final state. The probability for the production of like-sign
and opposite-sign leptons is equal, and the characteristics
of the two classes of final states are identical.  Observation of
this distinctive result would be extremely helpful in
identifying the origin of the events.

To be specific, we have taken
the branching ratio for gluino decay to the lightest chargino to be
$\br(\gl\to q\bar q^{\prime}\cpmone)= 0.58$, a result which
holds to good accuracy for all $|\mu|\gsim\mgl/3$, $m_Z$.\refmark\bgh\
Moreover, this value for the branching ratio is approximately valid for
nearly all MSSM parameters of relevance to the Tevatron gluino search.%
\foot{For example,
when $\mgl\lsim 200$ GeV and $\tanb\gsim 4$,
$\br(\gl\to q\bar q^{\prime}\cpmone)$ varies between about 45\% and 65\%
as $\mu$ is varied.}
If the $\slep$ and $\snu$ are heavier than the chargino,
the chargino decays dominantly
into the LSP plus a real (or virtual) $W^\pm$
which then decays $22\%$ of the time into electrons and muons.
Thus the branching ratio
for the decay chain $\gl \rta q\bar q \ell^\pm \nu \widetilde
\chi_1^0$ is likely to be as large as $13\%$, with equal probability
to produce a lepton of either sign.  (For simplicity,
the $\tau$--lepton will be neglected from our considerations.)
Since gluinos are produced in pairs, the
number of dilepton final states resulting from the decay of
the two gluinos would be about $1.6\%$ of all $\gl\gl$
events, of which half would have a pair of like-sign leptons.
However, if $\mslep$ and/or $\msnu< \mcpmone$, and $\mcpmone< m_W
+\mcnone$,
then the two-body decays of the $\cpmone$ to $\slep \nu$
and/or $\snu \ell$ will have approximately 100\% branching ratio into
leptonic final states (including the $\tau$-lepton).
If we neglect $\tau$--leptons, we find
$\br(\gl \rta q\bar q \ell^\pm \nu \cnone)$ close to $40\%$.  A remarkable
$15\%$ of all $\gl\gl$ events would yield a dilepton final state.

Thus, we propose that the Tevatron search for events with hadronic
jets (two from each gluino), missing energy due to
the LSP and neutrinos in the final state, and a dilepton pair
which can come in one of the following like-sign combinations:
$e^\pm e^\pm$, $e^\pm \mu^\pm$, $\mu^\pm \mu^\pm$, and the
corresponding opposite-sign combinations.  The
events would be very similar to those arising from $t\bar t$ production
in which the leptons come from primary decays of the $t$ and $\bar t$.
Thus, distinguishing the source of opposite-sign events might be difficult.
Because the efficiency for tagging $b$-jets is low and because
some fraction of $\gl\,\gl$ events will contain $b$-jets from
a hard radiated gluon, $b$-jets may not be a useful tool for
separating $t\bar t$ from $\gl\,\gl$ events.
For this reason we will focus on like-sign dilepton final states for which
$t \bar t$ production yields a background only through $\bar t\rta \bar
b \ell^-\nu$ and $t\rta b X$, $b\rta c\ell^-\nu$ (or
the corresponding charge-conjugated decay chain). This background
would be quite small since the lepton from
the $b$ decay would very rarely be isolated.

We have evaluated the rates for the dilepton signal in gluino
pair production at
the Tevatron.  In order to roughly account
for realistic experimental conditions, we have employed
a parton-level Monte Carlo, which included resolution smearing but
no fragmentation, to model the $\gl\gl$ events\foot{A description
of a similar program to analyze the characteristics of
supersymmetric events at the CERN
$p\bar p$ collider can be found in ref.~\bhk.}
The surprisingly large potential for gluino discovery at the Tevatron
becomes apparent by giving the number of dilepton
(opposite- plus like-sign) events obtained in a 25$\pbi$ year.
For the case where the $\cpmone$
decays to $\slep\nu$ or $\ell\snu$, the net branching
ratio of 15\% quoted above yields roughly 1440, 476, 183, 79, 37, and 18
dilepton events (before cuts) for gluino masses of 100, 120, 140, 160,
180, and 200 GeV, respectively. Dilepton rates originating from
$\cpmone\to W^{(*)} \cnone$ decays are a factor of roughly 9 smaller.

\REF\leplimit{
A  summary of results from all LEP collaborations
can be found in ref.~\pdg.}
To estimate the rates for {\it detectable} dilepton
events at the Tevatron, we assume a trigger which requires that
the leading (secondary) lepton has $E_T> 15$ (10) GeV.
In addition, we require $|\eta|<2.5$ and isolation for both
leptons.\foot{An isolated lepton is defined to be one
that is separated by at least 0.3 units in $\Delta R\equiv
[(\Delta\eta)^2+(\Delta\phi)^2]^{1/2}$ from any parton (or ``merged
parton jet'' if two or more partons are within 0.7 units in
$\Delta R$ of each other).}
Associated hadronic jets are not required.
The probability that events pass these cuts depends in
detail upon the masses of the particles involved in the decay
chains, but is roughly 50\% for much of parameter space.
We have explored a range of supersymmetric parameters
for which the mass $\mcpmone$ varies
between about 45~GeV (its current lower bound from
LEP data\refmark\leplimit) and 80~GeV,  while letting the
gluino mass vary between 100 and 200 GeV. In the corresponding
range of MSSM parameter space, the mass
of the LSP ($\cnone$) is approximately given by $\mgl/6$.

For $\cpmone\rta {\wpm}^*\cnone$, the 1.6\% net branching ratio
quoted earlier yields between 70 and 1 dilepton events
(of which half are like-sign) per 25 pb$^{-1}$ at the Tevatron for
$\mgl$ between 100 and 160 GeV.   In contrast, if the $\cpmone$ decays to
$\slep\nu$ and/or $\ell\snu$, the larger 15\% net branching ratio can
result in up to 1100 (15) dilepton events for $\mgl=100$ (200) GeV,
depending on the decay mode and the various masses.

\TABLE\tevatron{}
\topinsert
\vbox{ \leftskip=9pt \rightskip=\leftskip
 \ninepoint\baselineskip=11pt
\noindent
{\bf Table~\tevatron}.
Number of $\ell^\pm\ell^\pm$ plus $\ell^\pm\ell^\mp$
events after lepton cuts for various $\mgl$ values at the Tevatron with
an integrated luminosity of
25$\pbi$. The various decay modes of
the $\cpmone$ are indicated by $W$ ($W\cnone$), $\slep$ ($\slep \nu$),
$\snu$ ($\snu\ell$). These rates assume the branching ratios
quoted earlier (which are, in fact, typical over a wide range of
parameters).}
\vskip9pt

\thicksize=0pt
\tablewidth=\hsize
\hrule \vskip2pt \hrule
\begintable
Mode | $W$ & $W$ & $W$ & $\slep$ & $\snu$  & $\snu$  & $\slep$  & $\slep$
 & $\slep$ &  $\snu$ &  $\snu$ &  $\snu$ &  $\slep$ &  $\snu$ \cr
$M(\gl)$\hf| 140 & 140 & 140 & 160 & 160 & 160 & 180 & 180 & 180 & 180
& 180 & 180 & 200 & 200 \cr
$M(\cpmone)$\hf| 80 & 60 & 45 & 60 & 80 & 60 & 80 & 80 & 45 & 80 & 80 &
 60 & 80 & 80 \cr
$M(\slep~{\rm or}~\snu)$\quad\hf| $-$ & $-$ & $-$ & 40 & 70 & 50 & 75
& 55 & 40 & 40 & 60 & 50 & 75 & 40 \cr
 Events | 9 & 6 & 3 & 22 & 9 & 12 & 30 & 22 &
9 & 28 & 20 & 7 & 15 & 15 \endtable
\hrule \vskip2pt \hrule
\vskip9pt
\endinsert

To illustrate, we present in Table~\tevatron\ the
dilepton event rates (for an integrated luminosity of $L=25\pbi$)
for various cases which yield $\lsim 30$ events.
For comparison, with our cuts, 12 events are expected
{}from $t\bar t$ production for $\mt=140$ GeV, but all of these have
opposite-sign leptons. Thus,
for comparable gluino and top quark masses, the additional requirement of
two isolated {\it like-sign} leptons  would reduce the top quark rate
to a level far below that from gluinos.
However, a top quark significantly lighter than the
gluino might require that additional cuts be made to separate
the two signals.

The event rate depends strongly on the lepton cuts. The lepton spectrum
itself is quite sensitive to decay modes and decay product masses.
For the chargino three-body decay to $\cnone\ell\nu$ via a virtual $W$,
the lepton spectrum depends primarily on $\mcpmone-\mcnone$.
For chargino decays to $\slep \nu$ ($\snu\ell$),
the spectrum is essentially determined by
$\mslep-\mcnone$ ($\mcpmone-\msnu$).  Because of
the current experimental limit of
$\mslep>44$ GeV\refmark\leplimit,  most leptons from $\slep$ decay
pass our cuts for the $\mcnone$ values employed.
Thus, even for the relatively large gluino masses of 160 and 180 GeV,
the $\slep\nu$ decay event rates illustrated
are large enough to be in possible conflict with observed
rates at the Tevatron. In contrast, as illustrated in Table~\tevatron,
the event rate associated with $\cpmone\rta \ell\snu$
decays could be very small
since small values of $\mcpmone-\msnu$ are possible, leading to a soft
lepton that is unlikely to pass our cuts.
However, even for $\mgl=180$~GeV, if the $\snu$
mass is not close to $\mcpmone$ then the event rate
for the $\ell\snu$ mode is large.

Thus, if very few or no like-sign
dilepton events are found after accumulating $L=25\pbi$, then
improved limits on the gluino mass (as a function of other MSSM
parameters) will be attainable.  If $\cpmone\rta {\wpm}^*\cnone$ is the
dominant decay, a modest improvement of $\mgl>120$~GeV is possible, based
on the like-sign dilepton search.  In contrast,
if $\ell\snu$ or $\slep\nu$ decays of the $\cpmone$ are dominant
and the $\ell$ spectra are not suppressed by a small mass difference,
then  limits of order $\mgl\gsim 200$ GeV will be obtained
over the large region of parameter space for which
$\br(\gl\rta q\bar q^\prime \cpmone)$ is substantial.

Assuming that one has succeeded in isolating gluino candidates, it
is important to ask if one can estimate the
mass of the gluino, and the masses
of the decay products.
If we define a hadronic jet as having $E_T>15$ GeV and
pseudorapidity $|\eta|<2.5$,
the gluino pair events will typically have between 2 and 4 jets.
The $E_T$ spectrum of the jets is completely determined by
$\mgl-\mcpmone$. For the extreme case of $\mgl=100$ GeV
and $\mcpmone=80$~GeV, all jets are too soft ($E_T< 15$ GeV) for
identification. In contrast,
the missing $E_T$ spectrum is not strongly dependent on the
mass splittings or chargino decay modes, at least
for gluino and chargino masses in the ranges considered here,
and is centered at about 50 GeV.
We have performed detailed studies and find that
the most useful distributions for mass determinations
are $\ptlmax$, the $E_T$ of the most energetic lepton and $\ptjmax$, the
transverse energy of the jet with largest $E_T$.
For any given decay chain, these can
be used to estimate the mass differences, and the overall event rate
can then be used to determine the absolute mass scale, provided
statistics are adequate.
However, with $L=25\pbi$, event rates for $W$-mediated decays of
the $\cpmone$ at the Tevatron are inadequate.  For $\slep\nu$ and/or
$\snu\ell$ decays, predicted event rates are
such that $\pm25$ GeV would be achievable for $\mgl\lsim 160$ GeV.
But, as already noted, such large event rates are probably inconsistent
with current observations at the Tevatron.
Scenarios consistent with current
Tevatron rates would require $L\gsim100\pbi$ to achieve a $\pm25$ GeV
estimate of $\mgl$. In this regard, it is important to note
that knowledge of the relative importance of
the $\slep\nu$ and $\snu\ell$ decays of the $\cpmone$ is, in principle,
not required in order to estimate $\mgl$.
The $\ptlmax$ spectrum will allow an estimate of how
many events per produced gluino have passed the lepton cuts, and the
overall event rate will then allow an estimate of $\mgl$.  The $\ptjmax$
spectrum would then allow us to estimate $\mcpmone$.
Of course, if $\mcpmone$ is known from other sources
(\eg\ LEP-II), $\mgl$ can be estimated from the $\ptjmax$ spectrum
alone, without relying on the absolute event rate predictions which
will have substantial systematic uncertainties.

The like-sign signature for $\gl\gl$ production provides a
powerful tool, both for discovering evidence for supersymmetry and
for estimating the gluino mass.
It is the Majorana nature of the gluino that yields this
striking signature.  At the Tevatron collider,
favorable assumptions concerning gluino cascade decay branching
fractions, gluino production could yield
as many as 1100 dilepton events (after significant lepton cuts)
in the current 25 pb$^{-1}$ run, of which half would be like-sign.
Since large numbers of events are not seen, many new constraints on the
masses of the $\gl$, $\cpmone$, $\cnone$, $\slep$ and $\snu$ will be
obtained. If the handful of dilepton events currently observed at the
Tevatron were due to $\gl\gl$ production, an integrated luminosity
of at least $L=100\pbi$ would be required to obtain a $\pm15\%$ estimate
of the gluino mass.

\chapter{Conclusions}

The Tevatron is expected to have accumulated nearly 100$\pbi$ of data
by the end of the 1993-94 run.  The data that will be collected by the
CDF and D0 Collaborations will be sensitive to gluino masses up to
200~GeV.  The detection of a gluino signal may occur through a variety
of techniques.  In addition to the classic missing transverse energy
signal, experimenters should search for isolated like-sign di-leptons
and isolated hard photons in events with jets and missing energy.
Detection of a positive signal in at least two of these three channels
could be critical for confirming a gluino interpretation of the detected
events.  In addition, with a signal in more than one channel, one can
begin to zero in on specific values of some of the MSSM parameters.

In the absence of any of the signals mentioned above, one would
conclude that gluinos (and probably squarks as well) are too heavy to
be produced at the Tevatron in the near future.  Although a modest
improvement of gluino mass limits could be achieved with the upgraded
Fermilab main injector (perhaps, detectable gluinos with
$\mgl<300$~GeV for $L=1000\pbi$), one will need the services of a
supercollider (either LHC or SSC) to definitively establish or rule
out the existence of the gluino of low-energy supersymmetry.

\vskip\chapterskip
\centerline{\bf Acknowledgements}
\vskip .1in

The invitation and hospitality of Pran Nath and his colleagues
is much appreciated.
I would also like to thank S. Mikamo for encouraging me to do the
computations presented in section 3.  In addition, I gratefully
acknowledge the fruitful collaboration with
Mike Barnett and Jack Gunion that produced much of the material
presented in sections 2 and 4. Finally, I thank the hospitality
of the Aspen Center for Physics where the written version of
this work was completed.
This work was supported in part by the Department of Energy
and in part by the Texas National Research Laboratory Commission
grant \#RGFY93-330.

\refout
\figout

\bye